\documentclass[11pt,a4paper,onecolumn,reqno]{amsart}
\usepackage[a4paper, margin=2.3cm, bmargin=3cm]{geometry}

\usepackage{graphicx,stfloats} \usepackage{color}
\usepackage{amsmath}
\usepackage{amsaddr}
\usepackage{bm, soul}
\usepackage{mathtools}
\usepackage[colorinlistoftodos,prependcaption,textsize=tiny]{todonotes}
\usepackage{braket}
\usepackage{hyperref}
\usepackage[square,comma,sort&compress, numbers]{natbib}
\usepackage{enumerate}
\usepackage[version=4]{mhchem}

\usepackage{placeins}

\renewcommand{\st}[1]{}
\usepackage{mathrsfs}
\usepackage{siunitx}
 
\usepackage{multirow}

\usepackage{etoolbox}
\patchcmd{\pprintMaketitle}
  {\hrule\vskip12pt}
 {\hrule\vskip12pt\ifvoid\extrainfobox\else\unvbox\extrainfobox\par\vskip12pt\fi}
 {}{}

\newsavebox\extrainfobox
\usepackage{nomencl}

\usepackage{caption}
\usepackage{array}
\newcommand{\PreserveBackslash}[1]{\let\temp=\\#1\let\\=\temp}
\newcolumntype{C}[1]{>{\PreserveBackslash\centering}p{#1}}
\newcolumntype{R}[1]{>{\PreserveBackslash\raggedleft}p{#1}}
\newcolumntype{L}[1]{>{\PreserveBackslash\raggedright}p{#1}}

\title{Numerical modeling of pulverized iron flames in a multidimensional hot counterflow burner}

\author[stfs]{Xu Wen$^{1,*}$, Arne Scholtissek$^{1}$, Jeroen van Oijen$^{2}$, Jeffrey Bergthorson$^{3}$, Christian Hasse$^{1}$}
\email{wen@stfs.tu-darmstadt.de} 
\address[]{$^1$Technical University of Darmstadt, Department of Mechanical Engineering, Simulation of reactive Thermo-Fluid Systems, Otto-Berndt-Stra{\ss}e 2, 64287 Darmstadt, Germany\\
$^2$Technical University of Darmstadt, Department of Mechanical Engineering, Reactive Flows and Diagnostics, Otto-Berndt-Stra{\ss}e 3, 64287 Darmstadt, Germany
$^3$Department of Mechanical Engineering, Eindhoven University of Technology, 5600 MB Eindhoven, the Netherlands
$^4$Department of Mechanical Engineering, McGill University, 817 Sherbrooke Street West, H3A0C3 Montreal, Canada
}

\begin{document}
\pagestyle{plain}

\maketitle

\begin{abstract} 
Pulverized iron flames stabilized in a multidimensional hot counterflow burner are simulated using a numerical model, which is extended from the state-of-the-art model developed by Hazenberg and van Oijen (PCI, 2021) considering unsteady effects. The results are compared to available experimental data (McRae et al., PCI, 2019), including particle image velocimetry measurements, a direct flame photo, the flow field velocity and the flame speed for different iron and oxygen concentrations. The comparison shows that the particle dynamics and flame shape can be reasonably well predicted. The flow field velocity and flame speed also show quantitative agreement between the simulation and the experiment. Based on the validated simulation results, the iron combustion characteristics, including the thermal structures and the multidimensional effects, are analyzed for different oxidizer environments. The analysis shows that the iron particles undergo a transition from kinetic-controlled regime (up to ignition) to a diffusion-controlled regime (burning) at the central axis for both environments with the particle temperature being higher than the gas temperature at the flame front, which is indicated by the Damk{\"o}hler number. For the hot counterflow burner, there exist multidimensional effects, i.e., the temperature and Damk{\"o}hler number change along the radial direction.
\end{abstract}




\printnomenclature

\section{Introduction} \addvspace{10pt}
\label{Sec:1}

Iron as an energy carrier has advantageous physicochemical properties in terms of transport, storage and energetic utilization \citep{bergthorson2018recyclable, debiagi2022iron}. Moreover, iron is the only candidate that predominantly burns in heterogeneous mode forming porous oxide shell \citep{bergthorson2015direct} at the conditions of stoichiometric equivalence ratio, ambient pressure and temperature, when comparing the different metal fuels and the corresponding different combustion modes. The iron oxide is a micro-sized solid particle with complex structure, which can be reduced using renewable energy, forming a clean carbon-free cycle \citep{bergthorson2015direct, bergthorson2018recyclable}.

Both numerical \citep{goroshin1996quenching, hazenberg2021structures, soo2015reaction, soo2018combustion, sun1998structure, sun2003concentration} and experimental \citep{mcrae2019stabilized, julien2015flame, tang2011modes, palevcka2019new, ning2021burn, huang2021detailed, gill2010combustion, sun1990combustion, sun2003concentration} studies of metal particle combustion have been conducted for various configurations and operating conditions. The flame speed, flame temperature and quenching distance of iron powders have been measured for different oxidizer environments (e.g., oxygen/nitrogen, oxygen/argon, etc.), particle size distributions, and gravity environments \citep{mcrae2019stabilized, julien2015flame, tang2011modes}. It was found that the iron particle can burn in either a diffusion-controlled or kinetically-controlled regime depending on the iron and oxygen concentrations. In a diffusion-controlled combustion regime, the surface reaction rate is much faster than the oxygen diffusion rate from the bulk gas to the particle surface, while in the kinetically-controlled regime, the surface reaction rate becomes the limiting factor \citep{mi2021quantitative}. Although the flame speed, as a fundamental parameter in combustion, can be accurately measured for gaseous fuels, it is not straightforward for particle suspensions. This is because the burning velocity for dust flames not only depends on the particle shape, the purity and the size distributions, but it is also more sensitive to the flow gradients, the flow configuration and particle loading/inter-particle spacing, compared to a gaseous flame \citep{julien2017flame}. Thus, various uncertainties exist for the measurement of the laminar burning velocity, although similar results (20\,cm/s $\sim$ 30\,cm/s) were obtained in different configurations, including Bunsen burner, spherically-expanding flames, counterflow burner and freely propagating flames in tubes (see Fig.~5 in \citep{julien2017flame}). 

With different levels of detail, numerical models have been developed and applied to simulate zero-dimensional or one-dimensional (1D) metal flames in generic configurations \citep{goroshin1996quenching, hazenberg2021structures, soo2015reaction,  soo2018combustion, sun1998structure, sun2003concentration}. Based on the assumption of diffusion-controlled regime, Goroshin et al.~\citep{goroshin1996quenching} developed a simple analytical model for a quasi 1D dust flame, in which the flame burning speeds are calculated with algebraic equations formulated for fuel-lean and fuel-rich mixtures, respectively. Later, Soo et al.~\citep{soo2015reaction, soo2018combustion} extended the analytical model, in which the heterogeneous reaction rate is governed by the competition between the oxygen diffusion rate and the surface reaction rate following the idea initially proposed by Frank-Kamenetskii \citep{frank2015diffusion}. In this extended model, the particle is assumed to react heterogeneously via a one-step Arrhenius-type surface reaction, with the oxide product removed from the particle surface automatically. Based on the thermophysical particle model proposed by Soo et al.~\citep{soo2015reaction, soo2018combustion}, Hazenberg and van Oijen \citep{hazenberg2021structures} further extended the model for iron combustion to arbitrary equivalence ratios with a consistent formulation of the mass and energy transport between the dispersed and gaseous phases. In addition, the gaseous phase properties were described more realistically dependent on the local temperature and gas composition. The extended model was applied in a quasi-steady formulation to simulate 1D steady flame propagation of iron dispersed in air. The iron flame structure, burning velocity and flame temperature were investigated as a function of equivalence ratio. They found that the maximum burning velocity is located at fuel-lean conditions for the diffusion-controlled combustion regime studied. 

Although significant progress has been made, the developed numerical model has rarely been applied for simulations of multidimensional iron flames in a multidimensional configuration. This work extends the model proposed in \citep{hazenberg2021structures} to consider unsteady effects and investigates a multidimensional iron counterflow burner experimentally studied by McRae et al.~\citep{mcrae2019stabilized} for different iron and oxygen concentrations. The simulation results are compared to the experimental data, including the particle image velocimetry (PIV) measurements, a direct flame photo, the flow field velocity and the flame speed. In addition, the iron combustion characteristics, including the thermal structures and the multidimensional effects, are investigated for different oxidizer environments based on the simulation results for a specific iron concentration.

\nomenclature[C]{PIV}{Particle image velocimetry}

The remainder of this paper is organized as follows. Section \ref{Sec:2} presents the modeling methods, including the gaseous phase and the particle phase governing equations. Experimental and numerical setups are presented in Section \ref{Sec:3}. Results and discussions of the comparisons against experimental data and the iron combustion characteristics are given in Section \ref{Sec:4}. Finally, the findings are summarized in the conclusions. 

\section{Modeling methods} \addvspace{10pt}
\label{Sec:2}

\subsection{Gaseous phase governing equations}
\label{Subsec:21}

The gaseous phase is described with an Eulerian method. For the 3D iron counterflow flame studied, the governing equations for mass, momentum, species mass fractions and total enthalpy are solved. While the governing equations for mass and momentum are the same as those for the general multiphase combustion \citep{poinsot2005theoretical}, balance equations for the species mass fractions $Y_k$ and total enthalpy $ H_e$ can be written as,
    \vspace{-1mm}
\begin{equation}\label{eq:Yi}
\dfrac{\partial \rho Y_k}{\partial t} + \dfrac{\rho u_j Y_k}{\partial x_j} = \dfrac{\partial}{\partial x_j} \left(\rho \mathscr{D} \dfrac{\partial Y_k}{\partial x_j} \right) + \dot{S}_{C,k}  
\end{equation} 
    \vspace{-2mm}
\begin{equation}\label{eq:He}
\dfrac{\partial \rho H_e}{\partial t} + \dfrac{\partial \rho u_j H_e }{\partial x_j} = \dfrac{\partial }{\partial x_j} \left(\rho \alpha \dfrac{\partial H_e}{\partial x_j} \right) + \dot{S}_{C,H_e}
\end{equation}
where $\rho$ is the gaseous phase density, $t$ the Eulerian time, $u_j$ and $x_j$ the gaseous phase velocity and the Cartesian coordinate in the $j^{th}$ direction, respectively, $\mathscr{D}$ the mass diffusion coefficient calculated assuming the Lewis number to be unity, i.e., $\mathscr{D} = \lambda / \left(\rho C_{p,g} \right) $, where $\lambda$ is the thermal conductivity, and $C_{p,g}$ the specific heat capacity of the gaseous phase. The parameter $\alpha$ is the thermal diffusivity, and the two-way coupling source terms $\dot{S}_{C,k}$ and $\dot{S}_{C,H_e}$ are calculated as,
    \vspace{-1.3mm}
\begin{equation}
\dot{S}_{C,k} = \dfrac{- \delta_{k,s} }{V_{\mathrm{cell}}} \sum_{i=1}^n \left( \dfrac{\mathrm{d} m_p }{\mathrm{d} t} \right)_i
\end{equation}
    \vspace{-3mm}
\begin{equation}
\dot{S}_{C,H_e} = \dfrac{-1}{V_{\mathrm{cell}}} \sum_{i=1}^n \left( \delta H_{e,i}^{con} - \delta H_{e,i}^{che}  \right) + \dot{S}_{g}^{rad}
\end{equation}
where $V_{\mathrm{cell}}$ is the volume of the cell where particles reside, $i$ the summation index looping all particles in the local cell, $\delta_{k,s}$ the Kronecker delta function, which equals to 1 for species involved in the iron reaction with index $s$, $\mathrm{d} m_p / \mathrm{d} t$ the mass transfer rate of the iron particles, which will be given in the next subsection. The parameter $\delta H_e^{con}$ is the convective heat transfer between the particle phase and gaseous phase, $\delta H_e^{con}$\,=\,$ A_d  \mathrm{Nu} k_g \left(T - T_p \right) /d_p $. Here, $ A_d $ is the total particle surface area, $\mathrm{Nu}$ the Nusselt number calculated with the Ranz-Marshall model \citep{ranz1952evaporation}. Note that the Ranz-Marshall model formulated based on the spherical shape and continuum assumptions may not be strictly valid for submicron-sized particles. A transition-regime transport model that combines continuum and free molecular regimes specifically for iron combustion should be formulated in future works. The parameter $k_g$ is the thermal conductivity of the gaseous phase, $d_p$ the particle diameter, which is calculated according to the mass and density of the fresh iron and iron oxide, and the formulation will be given in the following context. The parameters $T$ and $T_p$ are the gaseous temperature and particle temperature, respectively. The chemical enthalpy transfer due to the mass transfer during iron combustion $\delta H_e^{che}$ is calculated as, $\delta H_e^{che}$\,=\,$\sum_{k}  H_{e,k}  \mathrm{d} m_p / \mathrm{d} t $, where the summation index $k$ loops all species involved in the iron reaction, and $H_{e,k}$ is the total enthalpy of species $k$. The source term due to radiative heat transfer between the particle phase and gaseous phase $\dot{S}_{g}^{rad}$ is calculated as, $\dot{S}_{g}^{rad}$\,=\,$\alpha_g \left( \left\langle G \right\rangle - 4 \sigma \left\langle T \right\rangle^4 \right)$. Here, $\alpha_g$ is the absorption coefficient of the gray gas, $\sigma$ the Stefan-Boltzmann constant, and $\left\langle G \right\rangle$ the cell-mean incident radiation, which is calculated using an extended P-1 model \citep{cheng1964two} for characterizing the radiative heat transfer in solid fuel flames \citep{wen2018analysis}. The absorption coefficient is set to $0.075 \, \mathrm{m^{-1}}$ following coal combustion \citep{wen2018analysis}. In the counterflow burner studied, the iron particles are ignited by the burnt products of stoichiometric mixture/air. Due to the fact that the absorption coefficient is unknown for the considered iron combustion system the overall absorption coefficient is set to be equal to that in the hydrocarbon solid fuel combustion system. The parameters in the radiation model specifically for iron combustion should be further investigated in future works. Note that the gaseous phase reaction rate does not appear in Eq.~\eqref{eq:Yi} since the hot products produced from a methane/air mixture at stoichiometric condition are used to ignite the iron particles in the experiment studied \citep{mcrae2019stabilized}. The gaseous phase flame is inside the nozzle and no reactions are considered in the gaseous phase outside the nozzle.

\nomenclature[B]{\(\lambda\)}{Thermal conductivity, $\mathrm{W/(m \cdot K)}$}
\nomenclature[A]{\(Y_k\)}{Mass fraction of species $k$}
\nomenclature[B]{\(\rho\)}{Gaseous phase density, $\mathrm{kg/m^3}$}
\nomenclature[A]{\(t\)}{Eulerian time, $\mathrm{s}$}
\nomenclature[A]{\(u_j\)}{Gaseous phase velocity in the $j^{th}$ direction, $\mathrm{m/s}$}
\nomenclature[A]{\(x_j\)}{Cartesian coordinate, $\mathrm{m}$}
\nomenclature[A\(D\)]{$\mathscr{D}\)}{Mass diffusion coefficient, $\mathrm{m^2/s}$}
\nomenclature[B]{\(\alpha\)}{Thermal diffusivity coefficient, $\mathrm{m^2/s}$}
\nomenclature[A\(S\)]{\(\dot{S}_{C,k}\)}{Source term in mass fraction conservation equation, $\mathrm{kg/(m^3\cdot s )}$}
\nomenclature[A\(S\)]{\(\dot{S}_{C,H_e}\)}{Source term in enthalpy conservation equation, $\mathrm{J/(m^3\cdot s )}$}
\nomenclature[A]{\(V_{\mathrm{cell}}\)}{Cell volume, $\mathrm{m^3}$}
\nomenclature[B]{\(\delta_{k,s}\)}{Kronecker delta function}
\nomenclature[B]{\(\delta H_e^{con}\)}{Convective heat transfer between the particle phase and gaseous phase, $\mathrm{J/kg}$}
\nomenclature[A]{\(A_d\)}{Diffusive particle surface area, $\mathrm{m^2}$}
\nomenclature[A]{\(A_r\)}{Reactive particle surface area, $\mathrm{m^2}$}
\nomenclature[A\(Nu\)]{\(\mathrm{Nu}\)}{Nusselt number}
\nomenclature[A]{\(k_g\)}{Thermal conductivity of the gaseous phase, $\mathrm{W/(m \cdot K)}$}
\nomenclature[A]{\(d_p\)}{Particle diameter, $\mathrm{m}$}
\nomenclature[A]{\(T, T_p\)}{Gaseous and particle temperatures, respectively, $\mathrm{K}$}
\nomenclature[B]{\(\delta H_e^{che}\)}{Chemical enthalpy transfer due to the mass transfer, $\mathrm{J/kg}$}
\nomenclature[A]{\(H_{e,k}\)}{Total enthalpy of species $k$, $\mathrm{J/kg}$}
\nomenclature[A\(S\)]{\(\dot{S}_{g}^{rad}\)}{Radiative source term in the gaseous phase equation, $\mathrm{J/(m^3 \cdot s)}$}
\nomenclature[B]{\(\alpha_g\)}{Absorption coefficient of the gray gas, $\mathrm{1/m}$}
\nomenclature[B]{\(\sigma\)}{Stefan-Boltzmann constant, $\mathrm{5.67 \cdot 10^{−8} \, W/(m^2 \cdot K^4 ) }$}
\nomenclature[A\(G\)]{\(\left\langle G\right\rangle \)}{Cell-mean incident radiation, $\mathrm{W/m^2}$}

\subsection{Particle phase governing equations} 
\label{Subsec:22}

The individual particles are tracked with a Lagrangian method using the particle-source-in cell method \citep{crowe1977particle}. Established correlations for heat and mass transfer are adopted, which have been widely validated in multiphase combustion, e.g., coal, biomass or spray combustion. Only the drag and gravity forces are of importance and the particle velocity is calculated as,
\begin{equation}\label{eq:dupdt}
\dfrac{\mathrm{d} u_{p,i}}{\mathrm{d} t} = \dfrac{3 C_D \rho}{4 d_p \rho_p } \left(u_i - u_{p,i} \right) u_{rel} + g_i
\end{equation}
where $u_{p,i}$ is the particle velocity in the $i^{th}$ direction, $C_D$ the drag coefficient calculated based on the assumption of solid sphere \citep{wen2013mechanics}, $\rho_p$ the particle density calculated as, $\rho_p$\,=\,$6 m_p/\left(\pi d_p^3 \right)$. The density of the fresh iron $\rho_{p,\mathrm{Fe}}$ is set to be 7874\,kg/m$^3$. The parameter $u_{rel}$ the relative velocity magnitude between the gas and particle, i.e., $u_{rel}$\,=\,$|\overrightarrow{u} - \overrightarrow{u_p} |$, and $g_i$ the gravity acceleration in the $i^{th}$ direction. 

\nomenclature[A]{\(u_{p,i}\)}{Particle velocity in the $i^{th}$ direction, $\mathrm{m/s}$}
\nomenclature[A]{\(C_D\)}{Drag coefficient}
\nomenclature[B]{\(\rho_p\)}{Particle density, $\mathrm{kg/m^3}$}
\nomenclature[B]{\(\rho_{p, \mathrm{Fe}}\)}{Fresh iron particle density, $\mathrm{kg/m^3}$}
\nomenclature[A]{\(u_{rel}\)}{Relative velocity magnitude between the gas and particle, $\mathrm{m/s}$}
\nomenclature[A]{\(g_i\)}{Gravity acceleration in the $i^{th}$ direction, $\mathrm{m/s^2}$}

Different from the combustion of other solid fuels (coal, biomass, etc.) where the hydrocarbon fuels undergo pyrolysis and partial-oxidization, the iron particle does not devolatilize and the oxidized products remain solid. Iron combustion is assumed to be a purely heterogeneous reaction process, where the oxidizer is transported to the particle surface through a boundary layer and the chemical reaction happens at the particle surface with oxidized product formed around the unburnt fresh particle \citep{bergthorson2015direct}. Thus, both the oxidizer diffusion rate to the particle surface and the surface reaction rate of the iron particle limit the mass transfer rate. Following the original idea proposed by Frank-Kamenetskii \citep{frank2015diffusion}, the mass transfer rate of iron particle is described as,
\begin{equation}\label{eq:dmpdt}
\dfrac{\mathrm{d} m_p }{\mathrm{d} t } = A_d \rho Y_{\mathrm{O_2}} \dfrac{  A_d  \mathscr{D}_0  \cdot  A_r  \mathscr{R}_k}{ A_d \mathscr{D}_0 + A_r \mathscr{R}_k}
\end{equation}
where $A_r$ is the reactive particle surface area. The parameter $\mathscr{D}_0$ describes the oxygen diffusion from the bulk gas to the particle surface, which is calculated as, $\mathscr{D}_0$\,=\,$\mathrm{Sh} \mathscr{D}_{s,\mathrm{O_2}} / d_p $. Here, $\mathscr{D}_{s,\mathrm{O_2}}$ is the oxygen diffusivity at the particle film layer, which is calculated based on the unity Lewis number assumption. The parameter $\mathrm{Sh}$ is the Sherwood number. The parameter $\mathscr{R}_k$ is the surface reaction rate, which is described with a single-step, first-order Arrhenius reaction, $\mathscr{R}_k$\,=\,$\mathscr{R}_0 \mathrm{ exp }\left[ -E_a /(R T_p)  \right]$. Here, $\mathscr{R}_0$ and $E_a$ are the pre-exponential constant and activation energy, respectively, and $R$ the universal gas constant. 

\nomenclature[A\(D\)]{\(\mathscr{D}_0\)}{Oxygen diffusion rate from the bulk gas to the particle surface, $\mathrm{m/s}$}
\nomenclature[A\(D\)]{\(\mathscr{D}_{s,\mathrm{O_2}}\)}{Oxygen diffusivity at the particle film layer, $\mathrm{m^2/s}$}
\nomenclature[A\(Sh\)]{\(\mathrm{Sh}\)}{Sherwood number}
\nomenclature[A\(R\)]{\(\mathscr{R}_k\)}{Surface reaction rate, $\mathrm{m/s}$}
\nomenclature[A\(R\)]{\(\mathscr{R}_0\)}{Pre-exponential constant, $\mathrm{m/s}$}
\nomenclature[A]{\(E_a\)}{Activation energy, $\mathrm{J/kmol}$}
\nomenclature[A]{\(R\)}{Universal gas constant, 8314\,$\mathrm{J/(kmol \cdot K)}$}

To quantify the relative importance of the oxidizer diffusion rate and surface reaction rate, a normalized Damk\"ohler number $\mathrm{Da}^*$ is introduced, which is defined as, 
\begin{equation}\label{eq:DaNumber}
\mathrm{Da}^* =  \dfrac{  A_r \mathscr{R}_k }{  A_d \mathscr{D}_0 +  A_r  \mathscr{R}_k } 
\end{equation}
With the above definition, the overall iron reaction rate is limited by the oxidizer diffusion rate if the value of $\mathrm{Da}^*$ is close to unity (i.e., diffusion-controlled regime), while it is limited by the surface kinetics if the value of $\mathrm{Da}^*$ approaches zero (i.e., kinetically-controlled regime). 

In accordance with previous modeling approaches \citep{hazenberg2021structures, bergthorson2015direct, thijs2022resolved}, the iron oxide surrounding the iron core is assumed to be porous such that the oxygen can diffuse towards the unreacted iron surface without impediments. Based on this assumption, the reactive and diffusive particle surfaces are equal and can be calculated as, $A_r = A_d = \pi d_{p}^2$. In previous works \citep{sun1990combustion}, the iron oxide shell was also assumed to be in a liquid phase, and the oxygen diffusion into the oxide shell is rate-limiting. In particular, experimental studies that provide further insights in these near particle processes are required.

\nomenclature[A\(Da\)]{\(\mathrm{Da}^*\)}{Normalized Damk\"ohler number}

The interphase heat transfer occurs when the particle temperature and the surrounding gas temperature are different. The iron particles are heated up or cooled down through convection, radiation and heat release due to particle reaction. Specifically, the instantaneous particle temperature can be determined as,
\begin{equation}\label{eq:dTpdt}
 C_{p,p} \dfrac{\mathrm{d} m_p T_p }{\mathrm{d} t} = \dfrac{m_p C_{p,g} \mathrm{Nu} }{3 \mathrm{Pr}} \left(\dfrac{T - T_p}{\tau_p} \right) + \dot{S}_{p}^{rad} +  \delta Q \dfrac{1}{s} \dfrac{\mathrm{d} m_p }{\mathrm{d} t}  + H_{a, \mathrm{O_2}} \dfrac{\mathrm{d} m_p }{\mathrm{d} t} + C_{p,p} T_{ref} \dfrac{\mathrm{d} m_p }{\mathrm{d} t} 
\end{equation}
where $C_{p,p} $ is the specific heat capacity of the particle phase, $\mathrm{Pr}$, the Prandtl number, is set to 0.6, and $\tau_p$, the particle relaxation time, is defined as, $\tau_p$\,=\,$\rho_p d_p^2 /(18 \mu) $ with $\mu$ being the dynamic viscosity of the gas. The specific heat capacity of the unburnt iron particle $C_{p,p}^0$ is set to be 677\,$\mathrm{J/(kg \cdot K)}$ The radiative source term is calculated as, $\dot{S}_{p}^{rad}$\,=\,$\varepsilon_p  A_d  \left(0.25 \left\langle G \right\rangle - \sigma T_p^4 \right)$ with $\varepsilon_p$ being the particle emissivity, which is set to 0.85. In this work, the combustion enthalpy $\delta Q$ is adjusted to ensure that the stoichiometric flame temperature agrees with the experimental data given that the final iron oxide products are unclear. For the simulation studied, the value of combustion enthalpy for iron oxidation is set to be 3720\,kJ/kg, which is lower than the value of 4844\,kJ/kg \citep{huang2014investigation}. The combustion enthalpy is scaled to ensure the calculated stoichiometric flame temperature agrees with the experimental observation. In fact, this treatment is consistent with that in previous works for iron combustion with the same scaling factor being adopted \citep{thijs2022resolved, hazenberg2021structures}. The parameter $s$ is the stoichiometric ratio, which relates the mass change of fresh iron to that of burning particle, i.e., $\mathrm{d}m_{p,\mathrm{Fe}}/\mathrm{d} t = - 1/s \cdot \mathrm{d}m_{p}/\mathrm{d} t $. The parameter $H_{a,\mathrm{O_2}}$ is the total enthalpy of the reacting oxygen, which is calculated according to the local temperature, i.e., $H_{a,\mathrm{O_2}} = \int_{T_{ref}}^{T} C_{p,g,k} \mathrm{d} T + \delta H_{c,k}^{ref}  $, where $T_{ref}$ is the reference temperature, and $\delta H_{c,k}^{ref}$ the formation enthalpy of species $k$ at the reference temperature $T_{ref}$. The reference temperature is set to be 300\,K in the simulations. 

\nomenclature[A]{\(C_{p,p}\)}{Specific heat capacity of the particle, $\mathrm{J/(kg \cdot K)}$}
\nomenclature[A]{\(C_{p,p}^0\)}{Specific heat capacity of the unburnt particle, $\mathrm{J/(kg \cdot K)}$}
\nomenclature[A]{\(C_{p,g}\)}{Specific heat capacity of the gaseous phase, $\mathrm{J/(kg \cdot K)}$}
\nomenclature[A]{\(C_{p,g,k}\)}{Specific heat capacity of species $k$, $\mathrm{J/(kg \cdot K)}$}
\nomenclature[A\(Pr\)]{\(\mathrm{Pr}\)}{Prandtl number}
\nomenclature[B]{\(\tau_p\)}{Particle relaxation time, $\mathrm{s}$}
\nomenclature[B]{\(\mu\)}{Gaseous phase dynamic viscosity, $\mathrm{kg/(m \cdot s)} $}
\nomenclature[A\(S\)]{\(\dot{S}_{p}^{rad}\)}{Radiative source term in the particle equation, $\mathrm{J/s}$}
\nomenclature[B]{\(\varepsilon_p\)}{Particle emissivity}
\nomenclature[B]{\(\delta Q\)}{Combustion enthalpy, $\mathrm{J/kg}$}
\nomenclature[A]{\(s\)}{Stoichiometric ratio}
\nomenclature[A]{\(T_{ref}\)}{Reference temperature, $\mathrm{K}$}
\nomenclature[A]{\(H_{a,\mathrm{O_2}}\)}{Total enthalpy of the reacting oxygen, $\mathrm{J/kg}$}
\nomenclature[B]{\(\delta H_{c,k}^{ref}\)}{Formation enthalpy of species $k$ at the reference temperature $T_{ref}$, $\mathrm{J/kg}$}

\begin{table}[h!]
\caption{Parameters in the combustion model for iron combustion.}
    \vspace{0mm}
\centerline{\begin{tabular} {L{80mm} L{20mm} L{20mm}}
\hline 
    Parameters & Values & Units \\
\hline
    Specific heat capacity of fresh iron, $C_{p,p}^0$ & 677 & $\mathrm{J/(kg \cdot K)}$ \\
    Fresh iron temperature, $T^0$ & 300 & K  \\
    Fresh iron density, $\rho_{p,\mathrm{Fe}}$ & 7874 & $\mathrm{kg/m^3}$  \\
    Iron oxide density, $\rho_{p,\mathrm{FeO}}$ & 5745 & $\mathrm{kg/m^3}$  \\    
    Particle emissivity, $\varepsilon_p$ & 0.85 & -- \\
    Absorption coefficient, $\alpha_g$ & $0.075$ & $\mathrm{1/m}$ \\
    Combustion enthalpy, $\delta Q$ & $3.72\cdot 10^6$ & J/kg  \\
    Stoichiometric ratio, $s$ & 0.28 & --  \\
    Reference temperature, $T_{ref}$ & 300 & K  \\
    Pre-exponential constant, $\mathscr{R}_0$ & $7.5 \cdot 10^6$ & m/s  \\
    Activation energy, $E_a$ & $1.2 \cdot 10^8$ & J/kmol  \\
\hline 
\end{tabular}}
	\label{table:parameters}
\end{table}

In this work, we consider that iron oxidizes to FeO as in previous work \citep{hazenberg2021structures}, i.e., $\mathrm{Fe + 0.5 O_2 \rightarrow FeO}$. Based on this assumption, the pre-exponential constant $\mathscr{R}_0$ and activation energy $E_a$ are set to be $7.5 \cdot 10^6$\,m/s and $1.2 \cdot 10^8$\,J/kmol, respectively, which are consistent with those reported in the work of Hazenberg and van Oijen \citep{hazenberg2021structures}. These reaction rate parameters are selected to ensure that an iron particle auto-ignites in air at a temperature of 850\,K, which is within the range of reported experimental data summarized by Breiter et al.~\citep{breiter1977models}. It was found that the $\mathrm{Fe_2O_3}$ is formed only in the cooled burnt mixture with no influence on flame speed, while $\mathrm{Fe_3O_4}$ might also not be formed in the flame \citep{hazenberg2021structures}. The density of the iron oxide $\rho_{p,\mathrm{FeO}}$ is set to be 5745\,$\mathrm{kg/m^3}$. Thus, the particle diameter $d_p$ can be calculated according to the mass and density of the fresh iron and iron oxide by assuming that the particle is a sphere, 
\begin{equation}\label{eq:diameter}
d_p = \sqrt[3]{\dfrac{6 V_p}{\pi}} = \sqrt[3]{ \dfrac{ 6 \left[ m_{p, \mathrm{Fe}}/\rho_{p,\mathrm{Fe}} + \left( m_p - m_{p,\mathrm{Fe}} \right) / \rho_{p,\mathrm{FeO}} \right]  }{\pi} } 
\end{equation}
where $V_p$ is the volume of the particle.

\nomenclature[A]{\(V_p\)}{Particle volume, $\mathrm{m^3}$}
\nomenclature[B]{\(\rho_{p,\mathrm{FeO}}\)}{Iron oxide density, $\mathrm{kg/m^3}$}

\section{Experimental and numerical setups} \addvspace{10pt}
\label{Sec:3}

The proposed numerical model is applied to a hot counterflow burner experimentally investigated by McRae et al.~\citep{mcrae2019stabilized}. In the experiment, the iron particles are loaded into a piston, which is driven upwards by a mechanical actuator. The dispersed iron powder is seeded by using a concentric sonic air knife fed by oxygen-argon mixtures. The iron particles transported by the gas mixture pass a laminarizing tube of $60 \, \mathrm{cm}$ length before exiting the bottom nozzle with a diameter of $15 \, \mathrm{mm}$. The bottom nozzle is equipped with an argon coflow. From the upper nozzle, a stoichiometric methane/air flame is stabilized inside the nozzle, and the hot burnt products of the flame exit the top nozzle and meet with the gas-particle mixture issuing from the bottom nozzle. Due to heat losses to the wall, the temperature of the hot products at the exit of the upper nozzle is measured at around $1450 \, \mathrm{K}$, which is considered to be sufficient to ignite the iron particles without additional heat. The fresh iron particle size distribution is measured with a scanning electron microscope, and it is determined as: $d_{10}$\,=\,$0.7 \, \mu \mathrm{m}$, $d_{50}$\,=\,1.4$\, \mu \mathrm{m}$ and $d_{90}$\,=\,2.5\,$\mu \mathrm{m}$. Two distinct oxidizer environments are investigated in the experiment, with $30 \% \, \mathrm{O_2} / 70 \% \, \mathrm{Ar} $ and $40 \% \, \mathrm{O_2} / 60 \% \, \mathrm{Ar} $ in volume fraction, respectively. In the experiment, the iron particle concentration increases gradually until a flame is formed. At low particle concentrations, the particles react individually and a flame cannot be formed since the heat released from small particles cannot accumulate before cooling down. However, above a critical concentration, a flame is formed spontaneously when the heat release rate is larger than the heat losses \citep{mcrae2019stabilized}. Both oxidizer environments correspond to fuel-lean conditions for all the iron concentrations studied. In fact, for the $30 \% \, \mathrm{O_2} / 70 \% \, \mathrm{Ar} $ environment, the stoichiometric iron loading is calculated to be $1926 \, \mathrm{g/m^3}$ if the iron oxide is assumed to be FeO, and it is $1552 \, \mathrm{g/m^3}$ if $\mathrm{Fe_3O_4}$ is the iron oxide. The iron concentration is monitored using a laser-attenuation probe, and the motion of iron particles is tracked using PIV \citep{mcrae2019stabilized}.

In the simulation, the computational domain is a 3D cylindrical field with a diameter of $30 \, \mathrm{mm}$ and a length of $20 \, \mathrm{mm}$. The computational domain is discretized with $3.96 $ million cells, with local refinements in the main flow. Two different levels of mesh resolutions are evaluated, i.e., one is the same as in \citep{wen2018analysis}, while the other one is finer as presented in the present manuscript. For the finer mesh resolution, the cell length in the axial direction is uniform and equals to 0.2\,mm. In the radial direction, the cell length is uniform and equals to 0.075\,mm in the main flow, while it is slightly stretched in the coflow with the largest cell length being 0.4\,mm. In the circumferential direction, the cell length is uniformly distributed, and 240 grid points are utilized for discretization.\footnote{The meshes adopted in this work are available from the authors upon request.} We found that the flame speed does not show obvious quantitative differences using these two different mesh resolutions. The iron particles transported by the oxidizer are injected from the lower port, while the burnt products of stoichiometric mixture/air (i.e., $Y_{\mathrm{CO_2}}$\,=\,0.1385, $Y_{\mathrm{H_2O}}$\,=\,0.123 and $Y_{\mathrm{N_2}}$\,=\,0.7385) are injected from the upper port. The particle temperature $T^0$ and velocity are equal to those of the local fluid, which are $300 \, \mathrm{K}$ and the measured velocity, respectively. The particle positions are randomly distributed at the inlet, and the size distributions are set following the measurement \citep{mcrae2019stabilized}. Specifically, the general size distribution model \citep{generalBC} is employed, i.e., the probabilities of particle size below 0.7\,$\mu$m, 1.4\,$\mu$m and 2.5\,$\mu$m are set to be $10 \%$, $50\%$ and $90 \%$, respectively. The upper limit of the initial particle size is set to be 6\,$\mu$m. The particle diameter between each size interval is randomly generated at each time step. As the measurement of the iron concentration has uncertainties ($\pm 50 \, \mathrm{g/m^3}$ \citep{mcrae2019stabilized}), a wide range of iron concentration exists in the experiment. In the simulation, the iron mass concentrations are set to $350 \, \mathrm{g/m^3}$ and $450 \, \mathrm{g/m^3}$ to account for the lower and higher iron concentrations in the experiment, respectively. Equal-sized iron particles are grouped as computational particles using the particle-source-in cell method \citep{crowe1977particle}. The volume fraction of the iron suspension can be estimated as $\gamma = \sum_{i=1}^{N_{\mathrm{cell}}} \pi d_p^3/(6V_{\mathrm{cell}}) $, where $N_{\mathrm{cell}}$ is the number of particles in the local cell. For the higher iron particle concentration of $450 \, \mathrm{g/m^3}$, the maximum particle volume fraction in the domain is calculated to be $0.03\%$. This value is well below the criteria value of $0.4 \%$ reported by Rizk and Elghobashi \citep{rizk1989two}, above which the particle-particle interactions become non-negligible and the Lagrangian model based on the particle-source-in cell method is not strictly valid. After the simulations reach convergence, around $59.3$ million and $73.5 $ million computational particles remain in the computational domain for the cases of $350 \, \mathrm{g/m^3}$ and $450 \, \mathrm{g/m^3}$, respectively. 

\nomenclature[B]{\(\gamma\)}{Particle volume fraction}
\nomenclature[A]{\(N_{\mathrm{cell}}\)}{Number of particles in the local cell}
\nomenclature[A]{\(T^0\)}{Fresh iron temperature, K}

For iron combustion, it was recently reported that different regimes, namely the continuous and the discrete regime, can be observed \citep{goroshin1998effect, palevcka2019new, goroshin2011reaction, tang2012propagation}. The discrete regime refers to the condition that the flame propagation becomes sensitive to the particles' distribution in space, and is practically independent of the particle combustion time \citep{goroshin1998effect}. The focus of the current work is not to identify the different regimes occurring in the investigated cases. The proposed modeling approach, which is similar to that in \citep{hazenberg2021structures}, was not validated to distinguish between the two regimes. Future research should particularly look at this aspect both with dedicated experiments and simulations.

The governing equations for the gaseous phase are solved with a finite volume method using the PIMPLE algorithm in the open-source package OpenFOAM \citep{weller1998tensorial}. The employed flow solver is a direct extension of the flow solver for pulverized solid fuel combustion, which has been extensively validated by comparing against the experimental data \citep{wen2021flamelet, wen2017large}. The coal combustion submodels in the flow solver are replaced by the iron combustion submodels. Besides the momentum and enthalpy transport equations, four governing equations are solved for the species mass fractions, i.e., $Y_{\mathrm{CO_2}}$, $Y_{\mathrm{H_2O}}$, $Y_{\mathrm{O_2}}$ and $Y_{\mathrm{N_2}}$. The mass fraction of argon is obtained by subtracting the sum of the above four species from unity. The iron particles can also react with $\mathrm{H_2O}$ and $\mathrm{CO_2}$ at high temperatures \citep{julien2015flame}, which is however not well understood and thus not considered in the simulations. For spatial derivatives, a second-order central differencing scheme is used, while for temporal derivatives, the implicit Euler scheme is used. All the simulations run for $0.1 \, \mathrm{s}$ (around 20 flow through times) to get converged results, and for another $0.06 \, \mathrm{s}$ for time-averaging.

\section{Results and discussions} \addvspace{10pt}
\label{Sec:4}

\subsection{Comparisons against experimental data}
\label{Subsec:41}

\begin{figure}[h!]
    \centering
    \captionsetup[subfigure]{labelformat=empty} 
    \includegraphics[width=0.6\textwidth]{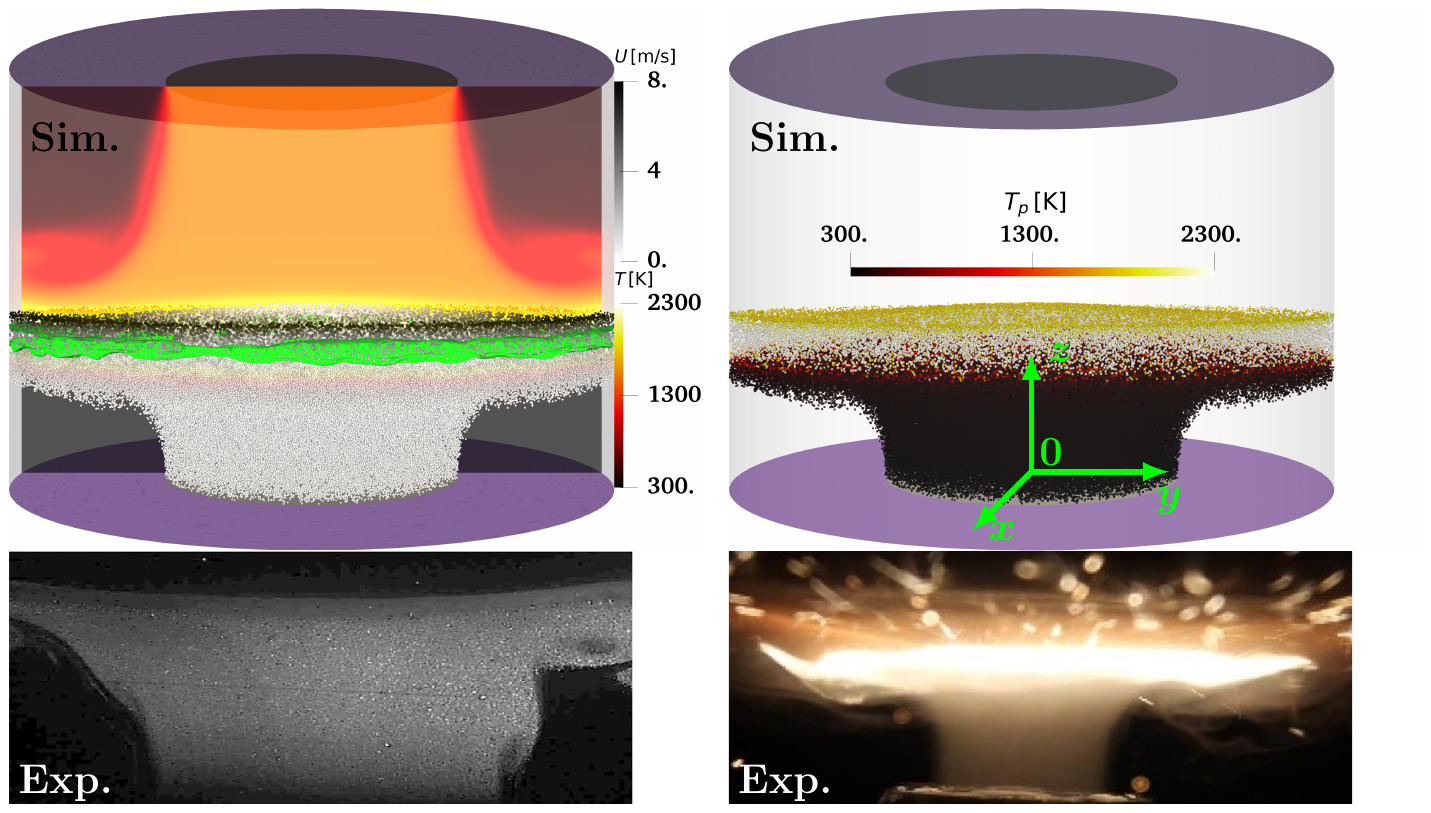}
    \caption{\label{fig:overall} Comparison with the PIV image (left) and the direct flame photo (right) \citep{mcrae2019stabilized} in the $40 \% \, \mathrm{O_2}$/$60 \% \, \mathrm{Ar}$ oxidizer environment with an iron particle mass concentration of 450 $\mathrm{g/m^3}$. The simulated particles are colored with their velocities (left) and temperatures (right). The predicted 2D snapshot of the gas temperature is also superimposed on the left figure. The green surface corresponds to the iso-surface of $T_{iso} = 2250 \, \mathrm{K}$, indicating the location of iron flame front.}
\end{figure}

The simulation results are first compared to the experimental images \citep{mcrae2019stabilized} to give an overall impression of iron combustion in a hot counterflow burner, as shown in Fig.~\ref{fig:overall}. The particles are colored with their velocities and compared to the PIV image (left), while those colored with the temperatures are compared to the direct flame photo (right) for the case in the $40 \% \, \mathrm{O_2}$/$60 \% \, \mathrm{Ar}$ oxidizer environment with a particle concentration of 450 $\mathrm{g/m^3}$. The predicted 2D snapshot of the gas temperature and the iso-surface of gas temperature of $T_{iso} = 2250 \, \mathrm{K}$ are also superimposed. It is seen that the overall particle flow pattern is similar to the PIV image. The direct flame photo shows that a flame is formed at the stagnation surface with bright radiating particles. In the simulation, the particles are oxidized at the corresponding location, and the surrounding gas is heated up by the strong heat release of burning iron particles, see also the next subsection. The hot surroundings in turn increase the reaction rate of all particles forming a flame front, as indicated by the green surface. The phenomenon that the thermal runway of a particle suspension takes place as a whole is due to self-heating of the mixture through the collective effect \citep{soo2015reaction, soo2018combustion, julien2015freely, cassel1964some}, which is similar to the combustion of homogeneous gaseous mixtures. The collective effect in a suspension describes a positive feedback mechanism, which is explained as follows \citep{rumanov1969critical}. In a suspension of reactive particles, the heat released from heterogeneous reactions is transferred from particles to the gas; in turn, the rising gas temperature reduces the heat transfer rate from the particles and accelerate the temperature increase of particles even further, thus, leading to a faster heterogeneous reaction rate due to its exponential dependence on temperature. However, we note that the collective effect depends on the availability of oxygen (diffusion limitation), see \citep{rumanov1969critical}. 

Note that although the snapshots shown in Fig.~\ref{fig:overall} appear overall symmetric, there are slight asymmetries due to a slightly non-ideal particle seeding in the experiment and the simulation, which eventually leads to locally varying particle number densities or residence times, which motivates a multidimensional configuration employed in the simulation. The multidimensional effects on the iron combustion characteristics will be investigated in detail in Section \ref{Subsubsec:422}.


\begin{figure}[h!]
    \centering
    \captionsetup[subfigure]{labelformat=empty} 
    \includegraphics[width=0.6\textwidth]{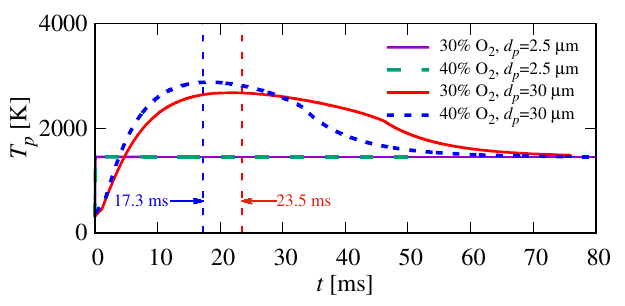}
    \caption{\label{fig:single} Time evolution of the particle temperature, which is obtained from the simulation of an individual particle of different sizes in $30 \% \, \mathrm{O_2}$ and $40 \% \, \mathrm{O_2}$ concentrations with a surrounding temperature of 1450\,K.}
\end{figure}

Without the collective effect, i.e., single particle combustion in a constant temperature atmosphere, the phenomenon is significantly different. Figure \ref{fig:single} shows the time evolution of the particle temperature, which is obtained for the simulation of an individual particle of different sizes in $30 \% \, \mathrm{O_2}$/$70 \% \, \mathrm{Ar}$ and $40 \% \, \mathrm{O_2}$/$60 \% \, \mathrm{Ar}$ environments with a gas temperature of 1450\,K. It is seen that with $d_p$\,=\,$d_{90}$\,=\,2.5\,$\mu\mathrm{m}$ (as in the experiment \citep{mcrae2019stabilized}), the particles are heated up to the ambient temperature rapidly without ignition occuring at all. Note that the 2.5\,$\mu\mathrm{m}$ iron particle does burn slowly in the kinetic regime with negligible Damk\"ohler number. The particle temperature keeps constant, which means that the heat losses from the burning particle to the gas are larger compared to the case with multiple particles. This finding confirms that the ignition of the iron suspension shown in Fig.~\ref{fig:overall} is most likely tied to the collective effect, consistent with \citep{soo2018combustion}. In contrast, for large particles with $d_p$\,=\,30\,$\mu\mathrm{m}$, the particles can be ignited in both environments, and the particle temperature reaches the peak value earlier in the $40 \% \, \mathrm{O_2}$ environment. This is expected since the mass transfer rate is proportional to $Y_{\mathrm{O_2}}$ in the diffusion-controlled regime, i.e., $\mathrm{d} m_p / \mathrm{d} t$\,$\propto$\,$  A_d  \rho Y_{\mathrm{O_2}} \mathscr{D}_0 $. In addition, the time to the maximum temperature is roughly proportional to the reciprocal of oxygen concentration, and the total time of the high temperature phase is close in different environments. This is consistent with the experimental observations for single iron particle combustion \citep{ning2021burn}. 

Note that the critical particle size that separates combustion modes (kinetic-controlled or diffusion-controlled) of individual particles cannot be determined accurately by the proposed numerical model since the effect of a growing oxide layer on the iron particle surface is not considered. A more sophisticated model for iron oxidation considering Fe-cation diffusion across the oxide layer is proposed in \citep{mi2021quantitative}, which will be considered in future simulations.

\subsection{Flame speed}
\label{Subsec:410}

\begin{figure}[!h]
    \centering
    \captionsetup[subfigure]{labelformat=empty} 
    \includegraphics[width=0.5\textwidth]{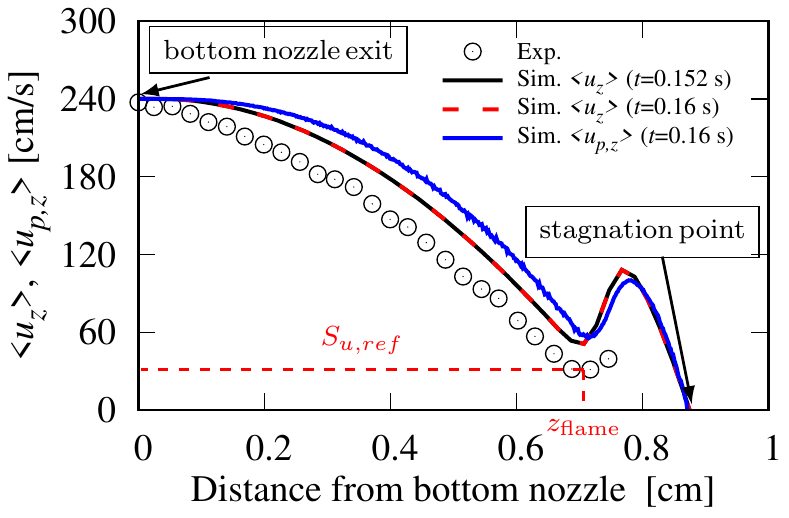}
    \caption{\label{fig:flowVelocity} Comparison of the measured velocity with the predicted gas and particle velocities averaged for different time durations. The oxidizer environment corresponds to $40 \% \, \mathrm{O_2}/ 60 \% \, \mathrm{Ar}$, while the iron concentration in the simulation is set to $450 \, \mathrm{g/m^3}$. The lowest flow velocity close to the fuel side corresponds to the flame speed $S_{u,ref}$, which is located at $z_{\mathrm{flame}} \, (\approx 0.7 \, \mathrm{cm}) $ in the experiment. }
\end{figure}

For the counterflow burner studied, the flame speed, denoted as $S_{u,ref}$, is defined as the minimum value of the flow velocity on the fuel side \citep{bergthorson2011experiments, mcrae2019stabilized}. Figure~\ref{fig:flowVelocity} compares the measured velocity with the predicted gas and particle velocities averaged for different time durations, along the central axis of the burner. The symbol $\left\langle \cdot \right\rangle $ indicates the time-averaged value. It is seen that the particle velocity closely follows the gas velocity, which justifies using iron particles as tracers in the PIV measurement \citep{mcrae2019stabilized}. In addition, both the particle velocity and the gas velocity profiles agree reasonably well with the measured velocity profile, with the flame speed being slightly over-predicted. Note that the increase of the flow velocity after the minimum value is due to the thermal expansion following iron particle ignition, which is also reproduced by the numerical model. 

\nomenclature[A]{\(S_{u,ref}\)}{Flame speed, $\mathrm{m/s}$}

\begin{figure}[!h]
    \centering
    \captionsetup[subfigure]{labelformat=empty} 
    \includegraphics[width=0.5\textwidth]{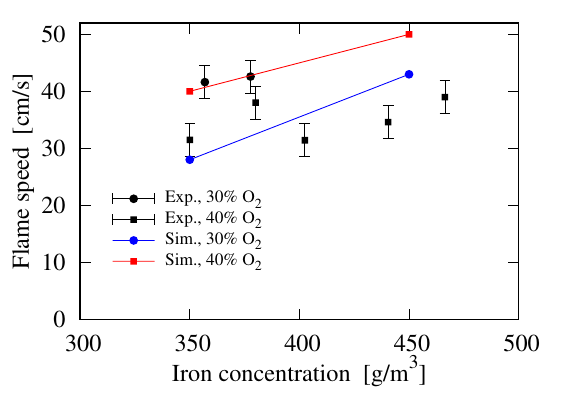}
    \caption{\label{fig:flameVelocity} Comparison of the predicted flame speed with the experiment for different iron and oxygen concentrations. The line connecting the predictions is just a guide to the eye.}
\end{figure}

Next, the flame speeds measured for different oxidizer and iron concentrations are compared to the simulation results in Fig.~\ref{fig:flameVelocity}. The error bar corresponds to the uncertainty in processing the PIV profiles to obtain the $S_{u,ref}$ value \citep{mcrae2019stabilized}. The measurement of iron concentration has a relatively high uncertainty, which is estimated to be $\pm 50 \, \mathrm{g/m^3}$ \citep{mcrae2019stabilized}. Although a quantitative comparison between the simulation and experiment is challenging, the measured flame speed is reasonably predicted. The remaining discrepancies could be attributed to the uncertainties in both experiments and simulations. Note that the iron combustion model is formulated based on several assumptions, e.g., the iron product is pure FeO, the iron particle reaction takes place in a heterogeneous mode without the potential vaporization being considered. From Fig.~\ref{fig:flameVelocity}, it can also be observed that in the simulation, the predicted flame speed increases as the iron concentration increases from $350 \, \mathrm{g/m^3}$ to $450 \, \mathrm{g/m^3}$, while the flame speed dependence on iron concentration is difficult to be identified in the experiment. In addition, the predicted flame speed increases for higher oxygen concentrations which suggests a diffusion-controlled combustion mode. The experimentally observed dependence of the flame speed on the oxygen concentration was attributed to the discrete wave propagation effect \citep{mcrae2019stabilized}. At fuel-lean conditions, the particles react faster than the heat diffusion from one particle to the others. Thus, the heat diffusion controls the propagation speed. Further investigations on the discrete wave propagation are required in future studies.

\subsection{Iron combustion characteristics}
\label{Subsec:42}

In this subsection, the iron combustion characteristics, including the thermal structures, multidimensional effects and the radiative heat transfer effects, in different oxidizer environments are analyzed. Here, only the case with an iron concentration of $450 \, \mathrm{g/m^3}$ is analyzed.

\subsubsection{Thermal structure analysis}
\label{Subsubsec:421}

\begin{figure}[h]
    \centering
    \captionsetup[subfigure]{labelformat=empty} 
    \includegraphics[width=1\textwidth]{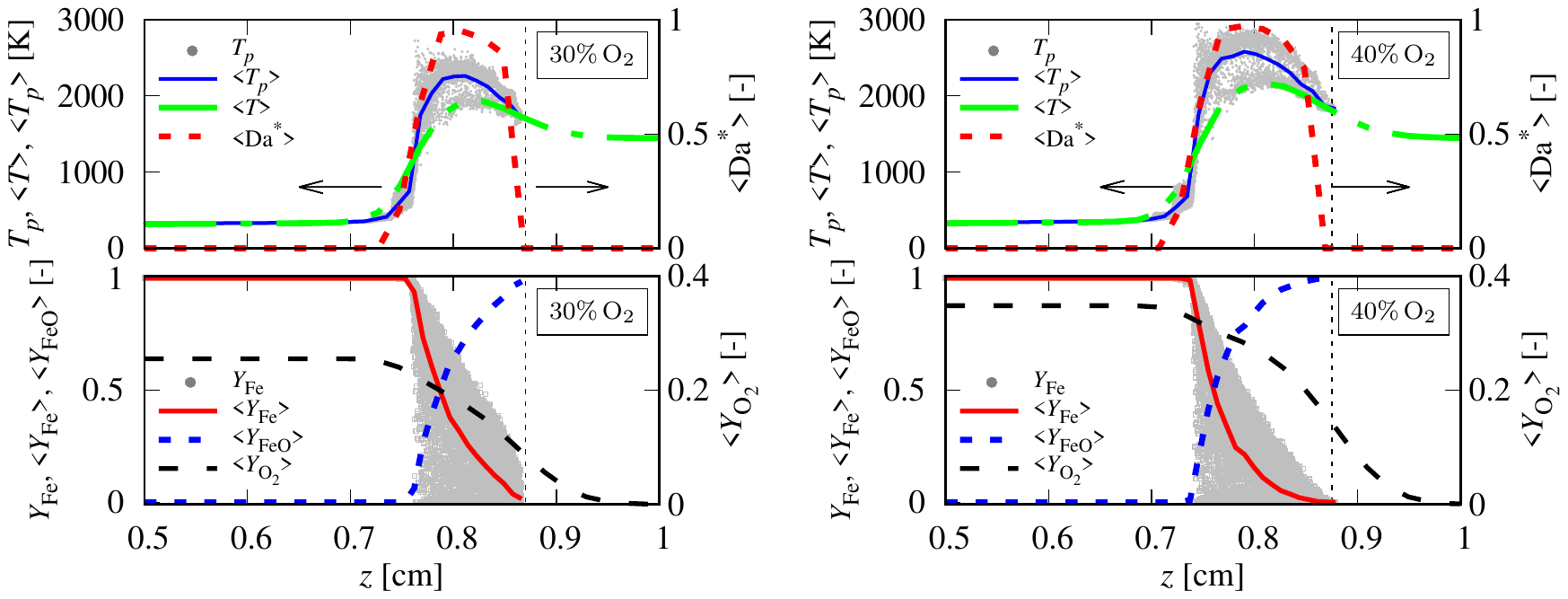}
    \caption{\label{fig:flamestructure35and40O2} Profiles of the thermal structure in $30 \% \, \mathrm{O_2}/ 60 \% \, \mathrm{Ar}$ (left) and $40 \% \, \mathrm{O_2}/ 60 \% \, \mathrm{Ar}$ (right) oxidizer environments for iron concentration of $450 \, \mathrm{g/m^3}$  along the central axis of the counterflow. Top: averaged gas and particle temperatures (including the instantaneous scatters), and the Damk{\"o}hler number; Bottom: mass fractions of unreacted iron $Y_{\mathrm{Fe}}$ (including the instantaneous scatters) and oxidized iron $Y_{\mathrm{FeO}}$, and the oxygen mass fraction. The vertical dashed line indicates the location of the stagnation point.}
\end{figure}

\begin{figure}[!ht]
    \centering
    \captionsetup[subfigure]{labelformat=empty} 
    \includegraphics[width=0.54\textwidth]{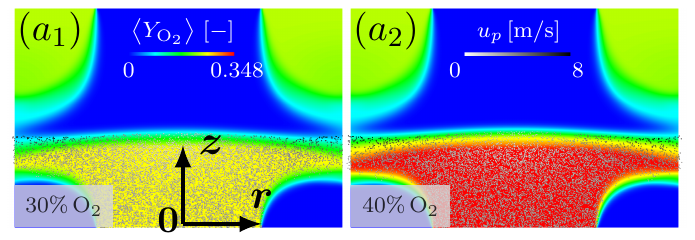}
    \caption{\label{fig:contourO2} Time-averaged contour plots of the $\mathrm{O_2}$ mass fraction for the ($a_1$) $30 \% \, \mathrm{O_2}$ and ($a_2$) $40 \% \, \mathrm{O_2}$ cases for an iron concentration of $450 \, \mathrm{g/m^3}$. The particles are superimposed and colored according to their velocity magnitude.}
\end{figure}

The thermal structures of the iron flames stabilized in the hot counterflow burner are analyzed for different oxygen concentrations. Figure~\ref{fig:flamestructure35and40O2} shows the thermal structure for the $30 \% \, \mathrm{O_2}$ (left column) and the $40 \% \, \mathrm{O_2}$ (right column) cases with an iron concentration of $450 \, \mathrm{g/m^3}$. The time-averaged gas and particle temperatures, and the Damk{\"o}hler number along the central axis of the counterflow are shown in the top row, while the mass fractions of unreacted iron $Y_{\mathrm{Fe}}$ and oxidized iron $Y_{\mathrm{FeO}}$, and the oxygen gas mass fraction are presented in the bottom row. The instantaneous particle temperature and the fraction of unreacted iron $Y_{\mathrm{Fe}}$ are shown as scatter, which are extracted from a cylindrical region with a diameter of $10 \, \mu\mathrm{m}$ around the central $z$-axis. The vertical dashed line indicates the location of the stagnation point. For the $30 \% \, \mathrm{O_2}$ case, the peak particle temperature is located upstream of the stagnation point and is higher than the local gas temperature, indicating a diffusion-controlled regime. The value of $\left\langle \mathrm{Da}^* \right\rangle $ in the corresponding region is close to unity, which verifies this statement. The distributions of $\left\langle Y_{\mathrm{Fe}} \right\rangle $ and $\left\langle Y_{\mathrm{FeO}} \right\rangle$ show that the fresh iron particles are completely oxidized before reaching the stagnation point, while the profile of $\left\langle Y_{\mathrm{O_2}} \right\rangle $ indicates a more smooth change due to the diffusivity of oxygen. The time-averaged contour plots of the $\mathrm{O_2}$ mass fraction for the two oxidizer environments are visualized in Fig.~\ref{fig:contourO2}. The particles are superimposed and colored according to their velocity magnitude. It can be observed that $\mathrm{O_2}$ changes smoothly around the flame front. In addition, it is seen that the particles remain in the region with non-negligible $\mathrm{O_2}$ concentration. 

Compared to the $30 \% \, \mathrm{O_2}$ case, the peak values of $\left\langle T \right\rangle $, $\left\langle T_p \right\rangle $ and $\left\langle \mathrm{Da}^* \right\rangle $ in the $40 \% \, \mathrm{O_2}$ environment are higher, as shown in the right column of Fig.~\ref{fig:flamestructure35and40O2}. The predicted variation trend of temperature in different oxidizer environments is consistent with the measured flame spectra and temperature, see Figs.~6 and 7 in ref.~\citep{mcrae2019stabilized}.  Note that with the pyrometric measurements of the reacting particles that burn as microflames, the experiment reported a higher flame temperature in the $40 \% \, \mathrm{O_2}$ case. However, the measured values cannot be compared to the simulation results in Fig.~\ref{fig:flamestructure35and40O2} directly, considering that the value in Fig.~\ref{fig:flamestructure35and40O2} is extracted only from the central axis. In fact, for the counterflow burner, multidimensional effects exist, i.e., the temperature changes significantly along the radial direction, which is investigated next.

\begin{figure}[!ht]
    \centering  
    \captionsetup[subfigure]{labelformat=empty} 
    \includegraphics[width=1\textwidth]{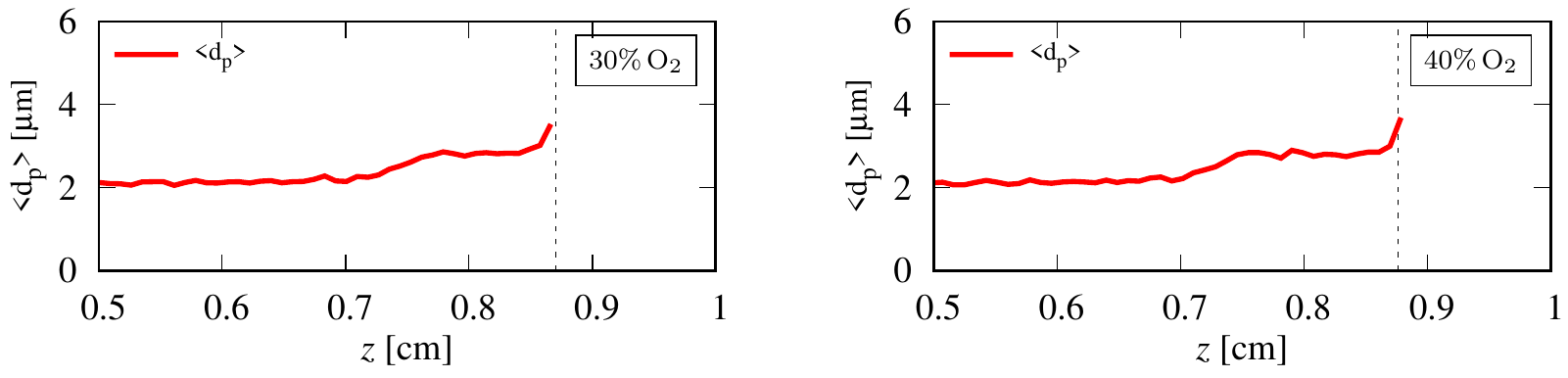}
    \caption{\label{fig:diameterZdirection} Mean particle diameter distribution in $30 \% \, \mathrm{O_2}/ 60 \% \, \mathrm{Ar}$ (left) and $40 \% \, \mathrm{O_2}/ 60 \% \, \mathrm{Ar}$ (right) oxidizer environments for iron concentration of $450 \, \mathrm{g/m^3}$  along the central axis of the counterflow. The vertical dashed line indicates the location of the stagnation point.}
\end{figure}

Finally, the mean particle diameter distribution along the central axis of the counterflow is shown in Fig.~\ref{fig:diameterZdirection} for the two oxidizer environments at an iron concentration of $450 \, \mathrm{g/m^3}$. The mean particle diameter is calculated by averaging the mass of particles in each bin. It can be observed that for both oxidizer environments, the particle size increases slightly at around $z=0.72$\,cm, which corresponds to the location where the iron oxidation takes place, see the profile of $\left\langle Y_{\mathrm{FeO}} \right\rangle $ in Fig.~\ref{fig:flamestructure35and40O2}. At the stagnation point, the particle size increases rapidly since the iron particles have been fully oxidized due to long residence times, i.e., the value of $\left\langle Y_{\mathrm{FeO}} \right\rangle $ equals to unity at this location, see Fig.~\ref{fig:flamestructure35and40O2}. The increase of the particle diameter towards the stagnation plane is due to the fact that the density of iron oxide is smaller than that of fresh iron (see Table \ref{table:parameters}). Comparing the two oxidizer environments shown in Fig.~\ref{fig:diameterZdirection}, the particle diameter distributions show no obvious differences.

\subsubsection{Multidimensional effects}
\label{Subsubsec:422}

\begin{figure}[h!]
    \centering
    \captionsetup[subfigure]{labelformat=empty} 
    \includegraphics[width=0.54\textwidth]{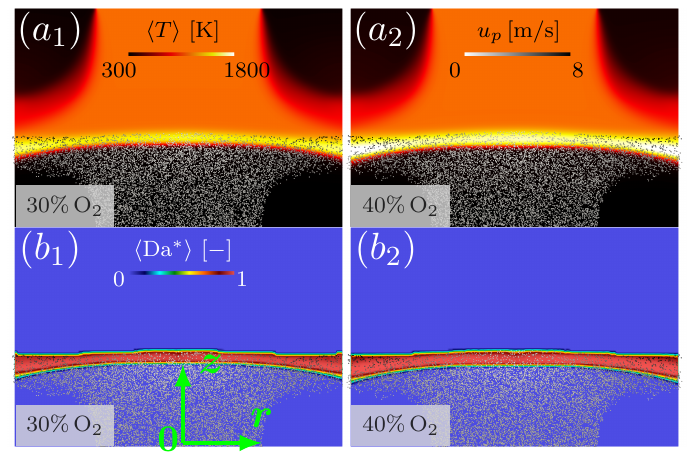}
    \caption{\label{fig:contours} Time-averaged contour plots of the ($a_1$,\,$a_2$) gas temperature, and ($b_1$,\,$b_2$) Damk{\"o}hler number for the $30 \% \, \mathrm{O_2}$ (left) and $40 \% \, \mathrm{O_2}$ (right) cases at the iron concentration of $450 \, \mathrm{g/m^3}$. The particles colored with their velocity magnitudes are superimposed.}
\end{figure}

The above analyses are all based on the central axis of the counterflow, which can be approximated by 1D simulations to a certain extent. However, for the multidimensional counterflow burner studied, the iron combustion characteristics are expected to change in the radial direction due to different particle residence times. To investigate this, Figs.~\ref{fig:contours}a and \ref{fig:contours}b show the time-averaged contour plots of the gas temperature $T$ and Damk{\"o}hler number $\mathrm{Da}^*$, respectively, for the $30 \% \, \mathrm{O_2}$ (left) and $40 \% \, \mathrm{O_2}$ (right) cases at an iron concentration of $450 \, \mathrm{g/m^3}$. The iron particles colored with the velocity magnitude are superimposed. It can be observed that for both oxygen concentrations, the flames become thicker in the radial direction due to the decreased strain rate of lower coflow velocity. The distribution of $\left\langle T \right\rangle$ is consistent with that of $\left\langle \mathrm{Da}^* \right\rangle $, i.e., the thicker the $\mathrm{Da}^*$ layer, the wider the high temperature region. This relation is an expected result since the iron particle mass transfer rate $\mathrm{d} m_p / \mathrm{d} t$ is proportional to the $\mathrm{Da}^*$ number, see Eq.~\eqref{eq:dmpdt}, which determines the heat release rate according to $\dot{Q}$\,=\,$ 1/s \cdot \delta Q \mathrm{d} m_p / \mathrm{d} t$. 

\nomenclature[A\(Q\)]{\(\dot{Q}\)}{Heat release rate, $\mathrm{J/s}$}

\begin{figure}[h!]
    \centering
    \captionsetup[subfigure]{labelformat=empty} 
    \includegraphics[width=0.6\textwidth]{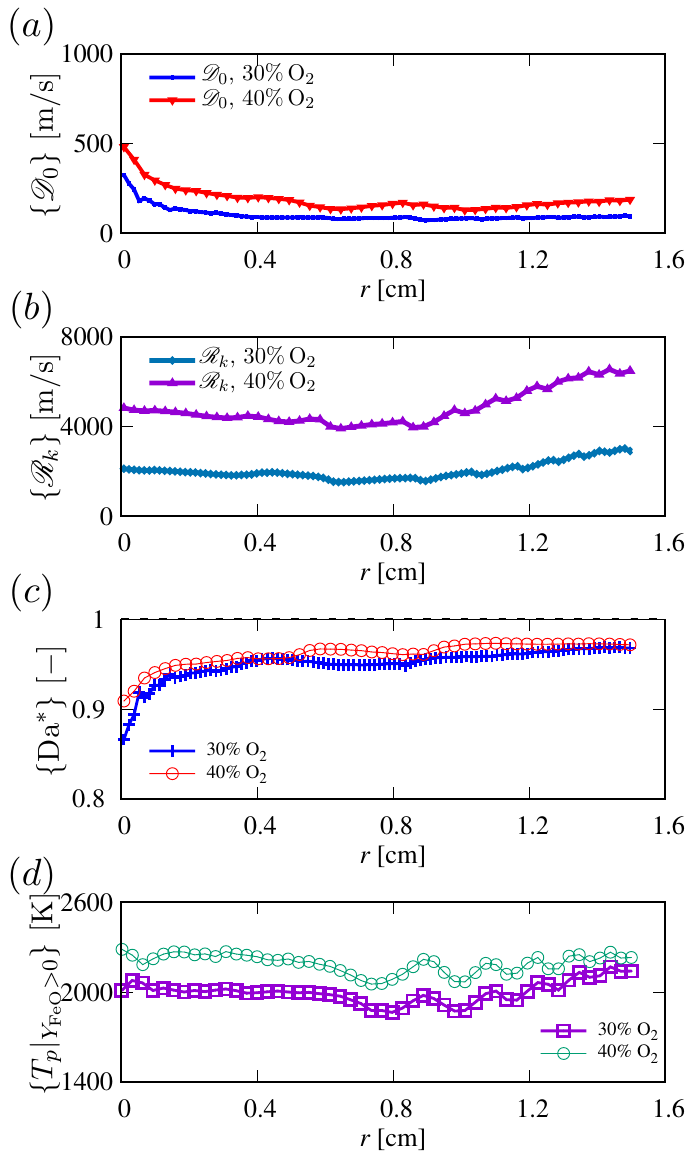}
    \caption{\label{fig:contours1} Radial profiles of (a) oxygen diffusion rate $\mathscr{D}_0$, (b) surface reaction rate $\mathscr{R}_k$, (c) Damk{\"o}hler number, and (d) burning particle temperature averaged for the $30 \% \, \mathrm{O_2}$ and $40 \% \, \mathrm{O_2}$ cases at the iron concentration of $450 \, \mathrm{g/m^3}$. The data is obtained from the entire domain along the radial direction.}
\end{figure}

As the $\mathrm{Da}^*$ value characterizes the competition between the oxygen diffusion coefficient from the bulk gas to the particle surface $\mathscr{D}_0$ and the surface reaction rate $\mathscr{R}_k$, it is important to investigate which process is dominant as the particles are transported outside. Figures \ref{fig:contours1}a and \ref{fig:contours1}b show the $\mathscr{D}_0$ and $\mathscr{R}_k$ values, respectively, averaged along the radial direction for both oxidizer environments, and the corresponding averaged $\mathrm{Da}^*$ value is shown in Fig.~\ref{fig:contours1}c. The symbol $\left\lbrace \cdot \right\rbrace$ indicates the spatially-averaged value, which is obtained from the entire domain averaged in the radial direction. It is seen that for both the $30 \% \, \mathrm{O_2}$ and $40 \% \, \mathrm{O_2}$ oxidizer environments, the $\left\lbrace \mathscr{R}_k \right\rbrace$ value is much larger than the $\left\lbrace \mathscr{D}_0 \right\rbrace$ value, which results in the corresponding $\left\lbrace \mathrm{Da}^* \right\rbrace$ value closer to unity, indicating a diffusion-controlled regime. In addition, it is seen that the value of $\left\lbrace \mathrm{Da}^* \right\rbrace$ increases from the central region with a value around 0.9 towards the outlet, which suggests an increased reaction rate as strain rate decreases. The diffusion coefficient $\mathscr{D}_0$ increases with the increasing of the oxygen concentration, as shown in Fig.~\ref{fig:contours1}a. This is an expected result since the oxygen concentration gradient at the particle film layer is higher in the $40 \% \, \mathrm{O_2}$ case than that in the $30 \% \, \mathrm{O_2}$ case.

Comparing the  $\mathscr{D}_0$ and $\mathscr{R}_k$ values shown in Figs.~\ref{fig:contours1}a and \ref{fig:contours1}b, it can be observed that the difference of $\left\lbrace \mathscr{R}_k \right\rbrace$ is much larger than that of $\left\lbrace \mathscr{D}_0 \right\rbrace$ as the oxygen concentration changes. The $\left\lbrace  \mathscr{R}_k \right\rbrace $ value in the $30 \% \, \mathrm{O_2}$ environment is much lower than that in $40 \% \, \mathrm{O_2}$. Recalling the definition of $\mathscr{R}_k$, i.e., $\mathscr{R}_k$\,=\,$\mathscr{R}_0 \mathrm{ exp }\left[-E_a /(R T_p) \right]$, the variation of $\mathscr{R}_k$ only depends on the local value of particle temperature $T_p$. As shown in Fig.~\ref{fig:contours1}d, the variation trend of the burning particle temperature $\left\lbrace T_p|_{Y_{\mathrm{FeO}} > 0} \right\rbrace$ is similar to that of $\left\lbrace \mathrm{Da}^* \right\rbrace$, which confirms the dominance of the diffusion-controlled regime for both cases. As $\mathscr{R}_k$ has an exponential relationship with $T_p$, a small difference of $T_p$ can change the $\mathscr{R}_k$ value significantly, which explains the much smaller value of $\mathscr{R}_k$ in the $30 \% \, \mathrm{O_2}$ case.

\begin{figure}[h!]
    \centering
    \captionsetup[subfigure]{labelformat=empty} 
    \includegraphics[width=0.6\textwidth]{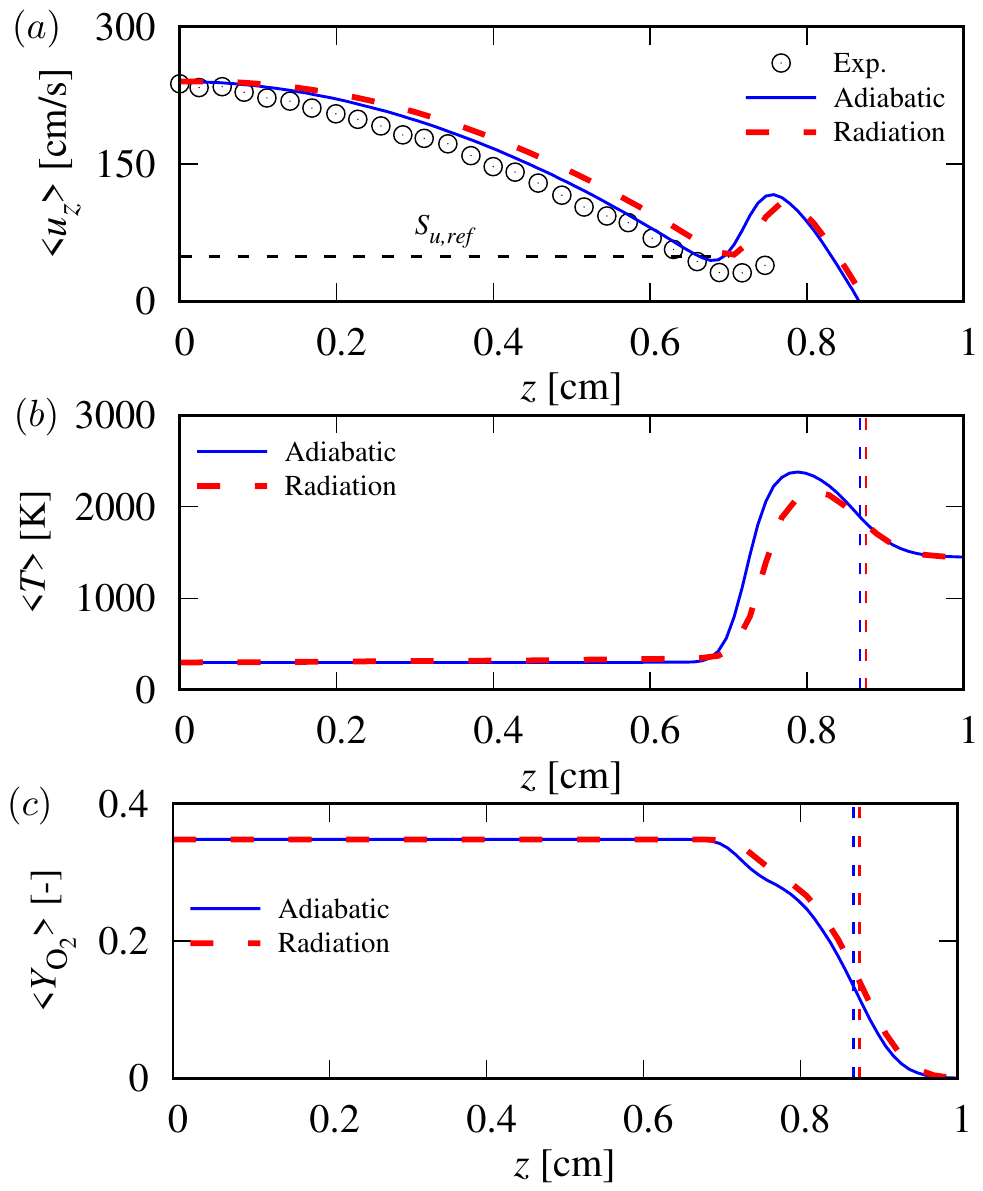}
    \caption{\label{fig:radiation} Radiative heat transfer effects on the velocities and the flame structures in $40\%$\,O$_2$/$60\%$\,Ar oxidizer environment for iron concentration of 450\,g/m$^3$. The horizontal dashed lines in (a) correspond to the flame speed $S_{u,ref}$, while the vertical dashed lines in (b, c) indicate the locations of the stagnation point.}
\end{figure}

\subsubsection{Radiative heat transfer effects}
\label{Subsubsec:423}

In this work, the iron combustion model presented by Hazenberg and van Oijen \citep{hazenberg2021structures} is extended by considering radiative heat transfer and how it affects the flame speed and the flame structure. To this end, an additional simulation is conducted for the case with $40\%$\,O$_2$/$60\%$\,Ar oxidizer environment and iron concentration of 450\,g/m$^3$, neglecting radiative heat transfer between the particle phase and the gaseous phase. Figure \ref{fig:radiation} shows the measured and predicted velocities, gas temperature and O$_2$ mass fraction for both the adiabatic and the radiative cases. The horizontal dashed line in Fig.~\ref{fig:radiation}a corresponds to the flame speed $S_{u,ref}$, while the vertical dashed lines in Figs.~\ref{fig:radiation}b and \ref{fig:radiation}c indicate the locations of the stagnation point. It can be observed that the radiation does not have obvious effects on the flame speed, although the flame position and the velocity profile are slightly changed. In Fig.~\ref{fig:radiation}b, the gas temperature distribution shows that the peak value of gas temperature in the adiabatic case is higher than that in the radiative case, which indicates non-negligible heat losses of hot burning particles. With radiation, the hot burning iron particles heat up the cool gases far from the particle, which leads to the temperature difference between the burning particles and the surrounding gas smaller. The smaller temperature difference results in the cooler gas temperature. The strong radiative heat transfer is associated with the fact that the radiative heat transfer is determined by the particle surface area, see the formulation of $\dot{S}_{p}^{rad}$ in Eq.~\eqref{eq:dTpdt}. The large total particle surface area of many small particles leads to the strong radiative heat transfer. In addition, it is seen that the position of the stagnation point in the non-adiabatic case moves towards the lower port. Due to the stronger reaction rate of higher gas temperature in the adiabatic case, the amount of O$_2$ in the adiabatic case is slightly lower than that in the non-adiabatic case, see the region between $z=0.7\,\mathrm{cm}$ and $z=0.9\,\mathrm{cm}$ in Fig.~\ref{fig:radiation}c.

\section{Conclusions} \addvspace{10pt}
\label{Sec:5}

An unsteady numerical model is proposed to simulate iron flames stabilized in a multidimensional counterflow burner operated at different conditions, in which the radiative heat transfer is considered. At first, the simulation results are compared to the available experimental data. The comparisons show that the predicted particle flow pattern and the flame shape are overall similar to the measurements. In addition, the flow field profile, together with the flame speed, is also reasonably predicted. The simulation results show that the flame speed increases as the oxygen concentration increases for the operating condition studied. Then, the iron combustion characteristics, including the thermal structures, the multidimensional effects and the radiative heat transfer effects, are analyzed for different oxidizer environments. The thermal structure analysis shows that the iron particles burn in a diffusion-controlled regime at the central axis for both oxidizer environments at the investigated iron concentration, with the particle temperature being higher than the gas temperature at the flame front, which is indicated by the Damk{\"o}hler number. For the counterflow burner, multidimensional effects exist, i.e., the temperature and Damk{\"o}hler number vary along the radial direction. For both cases, the diffusion-controlled particle combustion dominates in the radial direction, with the Damk{\"o}hler number being closer to unity moving towards the outlet. The trend of Damk{\"o}hler number is similar to that of the burning particle temperature for both oxidizer environments, which is consistent with a diffusion-controlled regime. The gas temperature is around 200\,K over-predicted when the radiative heat transfer between the iron particle phase and the gaseous phase is neglected. 


\section*{Acknowledgments} \addvspace{10pt}

This work was funded by the Hessian Ministry of Higher Education, Research, Science and the Arts - Clean Circles cluster project. Xu Wen acknowledges support through Marie Sk\l{}odowska-Curie Individual Fellowship (ID: 101025581) awarded by the European Commission under H2020-EU.1.3.2 scheme, and the fruitful discussions with Dr.~Xiaocheng Mi at Eindhoven University of Technology.








\bibliographystyle{unsrtnat_mod}
\bibliography{science}




\end{document}